\documentclass[showpacs,twocolumn,prb,superscriptaddress]{revtex4}
\usepackage{amssymb}

\usepackage{graphicx}
\usepackage{dcolumn}
\usepackage{bm}
\usepackage{epsfig}

\begin{document}

\title{Can one extract the electron-phonon-interaction from tunneling data
in case of the multigap superconductor MgB$_2$?}
\author{O. V. Dolgov}
\affiliation{Max-Planck Institut f\"{u}r Festk\"{o}rperforschung, Stuttgart, Germany}
\affiliation{Eberhard-Karls-Universit\"{a}t T\"{u}bingen, T\"{u}bingen,Germany}
\author{R. S. Gonnelli}
\author{G. A. Ummarino}
\affiliation{INFM-Dipartimento di Fisica, Politecnico di Torino,
10129 Torino, Italy}
\author{A. A. Golubov}
\affiliation{University of Twente, Faculty of Science and
Technology, 7500 AE Enschede, The Netherlands}
\author{S. V. Shulga}
\affiliation{Institute f\"{u}r Festk\"{o}rper- und Werkstofforschung, Dresden, Germany}
\author{J. Kortus}
\affiliation{Max-Planck Institut f\"{u}r Festk\"{o}rperforschung, Stuttgart, Germany}

\begin{abstract}
In the present work we calculate the tunneling density of states (DOS) of MgB%
$_{2}$ for different tunneling directions by directly solving the
two-band Eliashberg equations (EE) in the real-axis formulation.
This procedure reveals the fine structures of the DOS due to the
optical phonons. Then we show that the numeric inversion of the
standard \emph{single-band} EE (the only available method), when
applied to the \emph{two-band} DOS of MgB$_{2}$, may lead to wrong
estimates of the strength of certain phonon branches (e.g. the
$E_{2g}$) in the extracted electron-phonon spectral function
$\alpha^{2}F(\omega)$. The fine structures produced by the
two-band interaction at energies between 20 and 100 meV turn out
to be clearly observable only for tunneling along the $ab$ planes,
when the extracted $\alpha ^{2}F(\omega )$ contains the
combination $\alpha ^{2}F_{\sigma \sigma }(\omega
)$\textbf{+}$\alpha ^{2}F_{\sigma \pi }(\omega )$, together with a
minor $\alpha ^{2}F_{\pi \pi }(\omega )$\textbf{+}$\alpha
^{2}F_{\pi \sigma } (\omega )$ component. Only in this case it is
possible to extract information on the $\sigma$-band contribution
to the spectral functions. For any other tunneling direction, the
$\pi$-band contribution (which does not determine the
superconducting properties of MgB$_{2}$) is dominant and almost
coincides with the whole $\alpha^2F(\omega)$ for tunneling along
the $c$ axis. Our results are compared with recent experimental
tunneling and point-contact data.
\end{abstract}

\maketitle

There is a growing consensus that superconductivity in MgB$_2$
with a critical temperature $T_{c}\simeq 40K$ (Ref.
\onlinecite{Akimitsu}) is driven by the electron-phonon
interaction (EPI) (for a recent review see Ref. \onlinecite{MA}).
An important subject to address for a proper understanding of the
surprising physical properties of this material is the character
of the order parameter (or superconducting gap): is it constant
over the whole Fermi surface, or strongly
momentum dependent? The idea of multiband superconductivity in MgB$_{2}$ \cite%
{liu,sh,mazin,gol,br,Choi,mazinimp} is supported by many recent experimental
results from tunneling \cite{GiubileoPRL01,IavaronePRL02,Esk}, point contact %
\cite{SzaboPRL01,SchmidtPRL01,Gonnelli} and specific heat capacity
measurements \cite{BouquetPRL01}. These data directly support the
picture that the superconducting gap has two different values on
two qualitatively different parts of the Fermi surface, one
$\Delta _{\sigma }$ for the two quasi-two-dimensional $\sigma$
bands and another one $\Delta _{\pi }$ for the pair of 3D $\pi$
bands \cite{liu,mazin}.

While, within first-principles calculations of the electronic
structure and the EPI in this compound, there is an agreement
\cite{com1} on this qualitative picture, still disagreement is
present about the precise values of characteristic frequencies and
coupling constants. According to most calculations
\cite{liu,kong,bohnen,Choi}, the EPI or, equivalently, the
Eliashberg spectral function $\alpha ^{2}F(\omega )$ (EF) is
dominated by the optical boron bond-stretching $E_{2g}$ phonon
branch around 60 - 70 meV.

In principle, photoemission or optical measurements can also
deliver information on the EPI \cite{marsiglio}, although the main
experimental tool for the determination of the EPI in
superconductors so far is the tunneling measurement. This method
has been applied to standard superconductors with an isotropic
(constant in $\mathbf{k}$-space) superconducting gap and allowed
for the determination of the Eliashberg spectral functions in case
of many conventional low-temperature superconductors (see e.g.
Ref.~\onlinecite{wolf}).

The spectral EF is obtained from the first derivative of the tunneling
current
\begin{equation}
\frac{dI_{T}}{dV} \varpropto N_{T}=\left. \alpha
_{T}\text{Re}\frac{E}{\sqrt{ E^{2}-\Delta ^{2}(E)}}\right|
_{E=eV},  \label{dos}
\end{equation}%
where $V$ is the applied voltage, $\Delta (E)$ is the
\textit{complex} superconducting gap which depends on energy $E,$
and the factor $\alpha _{T}$ is determined by the properties of the tunneling barrier \textbf{\ }%
and the corresponding average of the Fermi velocities of
quasiparticles. The standard single-band procedure to obtain the
EF from the tunneling DOS can be found in textbooks
\cite{wolf,parks}. Another mathematically elegant method has been
proposed in Ref.\onlinecite{d1}. It has been used to investigate
conventional (low $T_{c}$) as well as high-temperature
superconductors \cite{nostro}.

Unfortunately, this approach is restricted to momentum independent
$s$-wave order parameters and cannot be used to describe
anisotropic superconductors as MgB$_2$. Nevertheless, there has
been a recent attempt to obtain the EPI in MgB$_{2}$ by using this
standard approach \cite{d1a}. The $E_{2g}$ phonon mode has been
resolved, but its predominance for the electron-phonon coupling
was questioned. More recently, the $E_{2g}$ mode has been also
resolved in point-contact spectra \cite{Naidyuk}.

The purpose of this paper is to clarify what information can be
extracted using this \emph{single-band} standard procedure if one
applies it to a \emph{two-band} superconductor. The starting point
is the theoretical study of the quasiparticle tunneling in
MgB$_2$-based junctions. The superconducting gap functions for the
$\sigma$ and $\pi$ band are obtained from an extended Eliashberg
formalism. The parameters for the two-band model utilized in this
work, which are based on first-principles electronic structure
calculations \cite{kong}, have been used before for a successful
description of specific heat \cite{gol} and tunneling \cite{br}
properties of MgB$_{2}$.
\begin{figure}[t]
\begin{center}
\includegraphics[width=\linewidth]{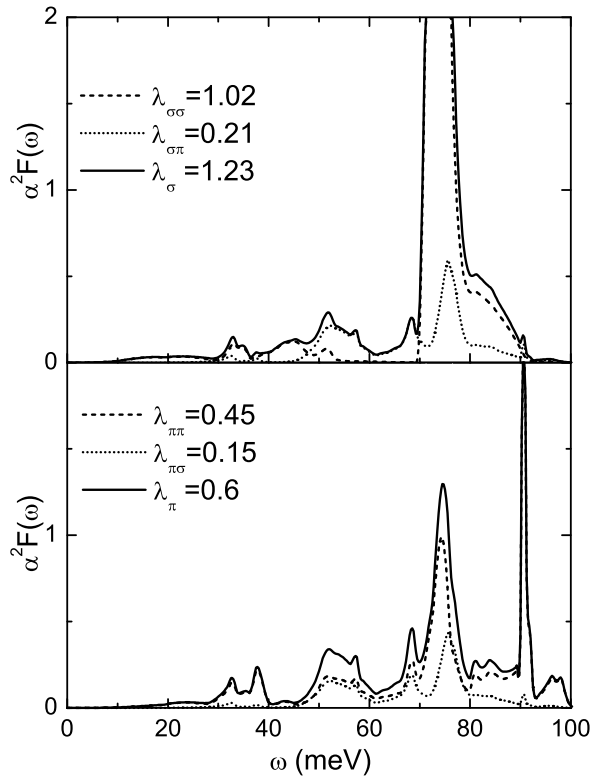}
\end{center}
\caption{The theoretical Eliashberg spectral functions of
MgB$_{2}$ for the two-band model used in this work (from
Ref.\onlinecite{gol}). The thick solid lines correspond to the
total $\protect\sigma $- and $\protect\pi $-band contributions. }
\label{fig1}
\end{figure}
The interband and intraband electron-phonon spectral functions
$\alpha ^{2}F_{ij}(\omega )$, where $i,j=\pi ,\sigma $ (see Figure
\ref{fig1}) and the Coulomb pseudopotential matrix $\mu
_{ij}^{\ast }$ (see Ref.\onlinecite{comment}) are the basic input
for the two-band Eliashberg theory. The theoretical conductance
curves of MgB$_{2}$ for different tunneling directions can be
obtained directly by solving the corresponding two-band Eliashberg
equations (EE) \cite{br} in the real-axis formulation. The only
free parameter is the normalization constant $\mu $ in the Coulomb
pseudopotential matrix which is fixed in order to reproduce the
experimental $T_{c}=39.4$ K.

One may see in Fig. \ref{fig1} that the $\sigma\sigma $ EPI is
dominated by the optical boron bond-stretching $E_{2g}$ phonon
mode. For other channels there are also important contributions
from low frequency modes (30-40 meV) and from high frequency
phonon modes ($\simeq$ 90 meV).
\begin{figure}[t]
\begin{center}
\includegraphics[width=0.5\textwidth]{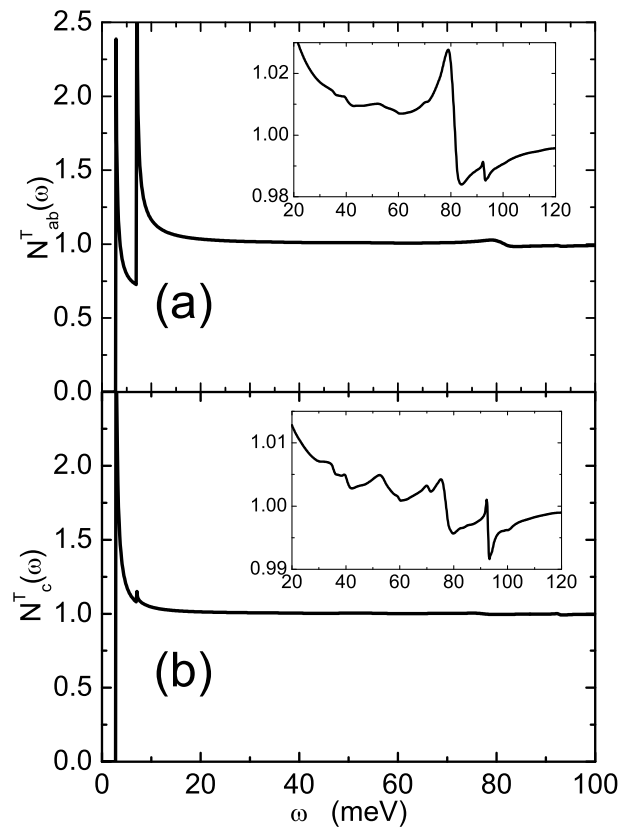}
\end{center}
\caption{(a) The calculated tunneling DOS in the
\textit{ab}-plane; (b) the calculated tunneling DOS along the
\textit{c}-axis. They are both obtained by the real-axis solution
of the two-band Eliashberg equations at T=0 K. The two insets show
the fine structures of the tunneling DOS due to the
electron-phonon interaction. } \label{fig2}
\end{figure}
\begin{figure}[t]
\begin{center}
\includegraphics[width=\linewidth]{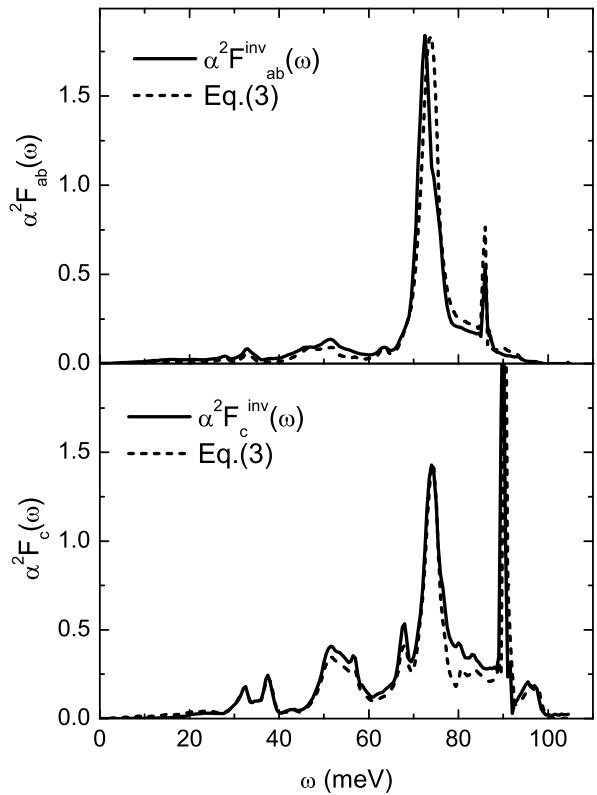}
\end{center}
\caption{The spectral Eliashberg functions along the
\textit{ab}-plane (upper panel) and along the \textit{c-}axis
(lower panel) obtained by inversion of the single-band Eliashberg
equations (solid line) applied to the tunneling DOS of Fig.2. The
dashed lines represent the least-square fits of these spectral
functions with two free parameters for $\protect\alpha^2
F_{\sigma}$ and $\protect\alpha^2 F_{\pi}$ as explained in the
text. } \label{fig3}
\end{figure}
In contrast to the case of a conventional junction described by Eq.(\ref{dos}%
), the conductance in a MgB$_{2}$-I-N tunnel junction is a
weighted sum of the contributions of the DOS of $\sigma $ and $\pi
$ bands, where the weights are determined by the corresponding
Fermi velocities (plasma frequencies) in the bands and by the
angle of the tunneling current with respect to the \textit{ab}
plane \cite{br}. Figure \ref{fig2} (a) and (b) shows the
calculated tunneling conductances in the \textit{ab}-plane and
along the \textit{c}-axis direction. In the case of MgB$_{2}$
according to Ref.\onlinecite{br},
\begin{eqnarray}
N_{ab}^{T}(\omega ) &=&0.33N_{\sigma }(\omega )+0.67N_{\pi }(\omega ),
\nonumber \\
N_{c}^{T}(\omega ) &=&0.01N_{\sigma }(\omega )+0.99N_{\pi }(\omega ),
\label{N}
\end{eqnarray}%
where $N_{\sigma}(\omega)$ and $N_{\pi}(\omega)$ are the partial
superconducting DOS.

The contribution of the $\pi $ band is always dominant even if
tunneling is almost in the \textit{ab} plane \cite{br}. In the
insets of Fig.2 the fine structures due to electron-phonon
interaction are shown. The maximum amplitude of these
structures is of the order of 0.5\% for measurements along the \textit{c}%
-axis and 2-3\% in the \textit{ab}-plane. The double-gap features
of the DOS visible in the \textit{ab}-plane $N_{ab}^{T}(\omega )$
should be experimentally observable even for a certain amount of
impurity scattering because the interband impurity scattering rate
appears to be very weak \cite{mazinimp}. On the other hand, due to
the smallness of the $\sigma $-band plasma frequency in the
\textit{c} direction, the conductance in the \textit{c}-axis
direction is almost totally determined by the $\pi $ band and
therefore no double-peak structure is expected in the conductance
spectrum. In fact this change in behavior has been experimentally
observed by spatially well-defined tunnel measurements over a step
edge in small MgB$_{2}$ single crystals \cite{IavaronePRL02} and
by directional point-contact spectroscopy in larger crystals
\cite{Gonnelli}.

In order to test the reliability of the numeric inversion
technique of the standard \emph{single-band} Eliashberg equation
when applied to a \emph{two-band} superconductor, we use the
calculated tunneling DOS shown in Fig. \ref{fig2} as an input. Of
course, in the case of a single-band superconductor the inversion
should reproduce the shape of the single-band Eliashberg function
used for the calculation of the tunneling DOS. This will not be
true anymore if one tries to invert a tunneling DOS which is
derived from a multiband Eliashberg theory. The inverted spectral
functions should correspond to the mixture of the $\sigma$- and
$\pi$-band contributions. We can introduce the following
functions:
\begin{eqnarray*}
\alpha ^{2}F_{\sigma }(\omega ) &=&\alpha ^{2}F_{\sigma \sigma }(\omega
)+\alpha ^{2}F_{\sigma \pi }(\omega ) \\
\alpha ^{2}F_{\pi }(\omega ) &=&\alpha ^{2}F_{\pi \pi }(\omega )+\alpha
^{2}F_{\pi \sigma }(\omega ).
\end{eqnarray*}%
Namely, these functions determine the normal state properties in
the two-band model. Tunneling measurements can only give
information on these combinations of $\alpha ^{2}F_{i,j}(\omega )$
($i,j = \sigma,\pi$) which are indicated by thick
lines in Fig.~\ref{fig1}. In order to illustrate this point we show in Figure \ref%
{fig3} (a) and (b) the results of the inversion of our calculated
tunneling DOS of Fig.2 (a) and (b) using a standard single-band
code (solid lines). The effect of the single-band EE inversion is
estimated by a least-square fit to the inverted spectral functions
with the weights for $\alpha ^{2}F_{\sigma }$ and $\alpha ^{2}F$
$_{\pi }$ as free parameters. These fits are shown by the dashed
curves in Fig.3. Since tunneling along the $c$-direction --due to
the very different plasma frequencies in the two bands \cite{br}--
practically corresponds to a single-band case, the inversion
properly reproduces the $\alpha ^{2}F_{\pi}$ as one expects. In
contrast, in the case of tunneling along the $ab$-direction (which
corresponds to an actual multiband situation) the weights are
quite different from the theoretical expectation. The results of
the fit are summarized below:
\begin{eqnarray}
\alpha ^{2}F_{ab}(\omega ) &\simeq &0.31\alpha ^{2}F_{\sigma }(\omega
)+0.16\alpha ^{2}F_{\pi }(\omega )  \nonumber \\
\alpha ^{2}F_{c}(\omega ) &\simeq &0.01\alpha ^{2}F_{\sigma }(\omega
)+0.99\alpha ^{2}F_{\pi }(\omega )  \label{sys}
\end{eqnarray}%
As one may see the $\sigma$-band spectral functions $\alpha
^{2}F_{\sigma \sigma }(\omega )+\alpha ^{2}F_{\sigma \pi }(\omega
)$ play an essential (and amplified) role only if the contribution
of $N_{\sigma}(\omega)$ is significant. The numerical simulations
show that about 33\% of the $\sigma$-band contribution in the
tunneling DOS corresponds to a contribution of about 66\% of
$\alpha ^{2}F_{\sigma }(\omega )$ in the
effective $\alpha ^{2}F_{ab}(\omega )$. These results are reasonable since $%
\lambda _{\pi }=(\lambda _{\pi \pi }+\lambda _{\pi \sigma })\approx 0.6$
while $\lambda _{\sigma }=(\lambda _{\sigma \sigma }+\lambda _{\sigma \pi
})\approx 1.23$. Somewhat simplifying it seems that the contributions of the
tunneling DOS to the phonon structures are weighted by the corresponding
coupling constants. The inversion of the \textit{c}-axis case results in a $%
\alpha ^{2}F_{c}(\omega )$ that is almost exactly the sum of the
$\alpha ^{2}F_{i}(\omega )$ components taken with the same weights
present in the sum of the corresponding superconducting $N_{\pi
}(\omega )$ and $N_{\sigma }(\omega )$ (see, Eqs.\ref{N}). In this
case, the coupling constant from the inversion is almost equal to
$\lambda _{\pi }$, and therefore the effective Eliashberg function
will show strong contributions from low and high frequency
phonons.

The above results show that from tunneling measurements at very
low temperatures and by numeric inversion of the standard
single-band Eliashberg equations one can obtain reliable
information on the EPI in MgB$_{2}$ only
for the $\pi $ band. For doing this we can use the Donetsk's inversion program %
\cite{d1} which allows to find the $\alpha ^{2}F(\omega )$ for a
single-band superconductor.

The interesting point is that measuring single crystals in the
clean limit one should see different phonon contributions for
different tunneling directions, in accordance with the two-band
model. From tunneling measurements exactly along \textit{ab} and
\textit{c} directions it is possible to extract information on $\alpha ^{2}F_{\pi}(\omega)$ and on $%
\alpha ^{2}F_{\sigma}(\omega),$ by solving the system of equations
(\ref{sys}) and this could be a useful method for testing the
two-band model and identifying the phonon modes responsible for
superconductivity in MgB$_{2}$. From the experimental point of
view this possibility could remain only virtual since the
smallness of the phonon structures expected in the tunneling along
the \emph{c} axis (see the inset of Fig.2 (b)) could prevent a
correct inversion of the EE in the presence of noise.

In polycrystalline samples the $\pi $ band dominates the tunneling current %
\cite{br} and it is only possible to extract information about the
combination of spectral functions $\alpha ^{2}F_{\pi }(\omega )$,
which does not play an important role for the superconducting
properties of MgB$_{2}$. The recent work of D'yachenko \textit{et
al.}, \cite{d1a} which claims the experimental determination of
$\alpha ^{2}F(\omega )$, is very likely contaminated by strong
contributions from the $\pi $ band. A value of the coupling
constant $\lambda =0.9$ was reported in Ref.\cite{d1a}.
Unfortunately, such data cannot give information on the nature of
superconductivity in a two-band superconductor as MgB$_{2}$ which
is driven by the interaction in the $\sigma $ band.

The inversion of the calculated \textit{ab}-plane conductance $%
N_{ab}^{T}(\omega )$ shows that a significant contribution to the
phonon structures comes from $\sigma $ bands and this is reflected
in the resulting $\alpha ^{2}F_{ab}(\omega )$. Therefore, only in
junctions with the tunneling current running along the $ab$ planes
and at low temperature one can observe the fine structures of the
superconducting DOS produced by the electron-phonon interaction at
energies between 20 and 100 meV. In recent point-contact
measurements \cite{Naidyuk} the anisotropic EPI was observed,
though no quantitative estimate was presented. Further experiments
on high-quality tunnel junctions are needed in order to obtain
data allowing for a quantitative estimate of the EPI, which should
also take into account the 'tunneling cone' effect \cite{wolf},
i.e. the distribution of tunneling
angles. According to our result shown in Eq.(\ref{sys}), separate studies of \textit{%
ab}-plane and \textit{c}-axis tunneling conductances may allow a
quantitative estimate of the EPI and should thus provide a crucial
test for the first-principle results of the two-band model.


\begin{thebibliography}{99}
\bibitem{com1} for a discussion why one should observe two gaps, but not a
gap distribution see Ref.\onlinecite{comment}.
\end{thebibliography}
\end{document}